\documentclass[12pt]{article}
\usepackage{times}
\usepackage{amssymb,latexsym}
\usepackage{amsmath}
\usepackage{graphicx}
\usepackage{graphics}
\usepackage{extarrows}
\usepackage{float}
\usepackage{cite}
\usepackage{epsfig}
\usepackage{subfig}
\usepackage{mathrsfs}
\usepackage{fancyhdr}
\usepackage{tensor}
\usepackage{multicol}
\usepackage{setspace}
\usepackage[usenames]{color}
\usepackage[left=2.5cm,top=3cm,bottom=2.5cm,right=2.5cm,columnsep=.6cm]{geometry}

\newcommand{\al}[1]{\begin{align}#1\end{align}}
\newcommand{\bs}{\begin{split}}
\newcommand{\es}{\end{split}}
\newcommand{\eqr}{\eqref}
\newcommand{\mc}{\mathcal}
\newcommand{\pa}[1]{\left(#1\right)}
\newcommand{\pb}[1]{\left[#1\right]}

\numberwithin{equation}{section}

\usepackage[colorlinks=true,linkcolor=blue,citecolor=blue,urlcolor=blue]{hyperref}
\usepackage[nameinlink]{cleveref}

\begin{document}

%--------------------------------------------------------------------------------------------------------------------------

\title{\vspace{-2cm} \textbf{Chern-Simons Splitting of 2+1D Pure\\ Yang-Mills Theory at Large Distances}}
\author{T. Yildirim${ }^{a,b}$\footnote{\href{mailto:tuna.yildirim@asu.edu}{tuna.yildirim@asu.edu}}
\vspace{.3cm}\\ 
\small\textit{${ }^a$Department of Physics and Astronomy, The University of Iowa, Iowa City, IA 52242} \\
\small\textit{${ }^b$Department of Physics, Arizona State University, Tempe, AZ 85287} \\
\vspace{-0.5cm}
\small\date{\normalsize\today}}

\maketitle
\vspace{-1.3cm}

%--------------------------------------------------------------------------------------------------------------------------

\begin{center}
\noindent\line(1,0){415}
\end{center}

\begin{abstract}

\noindent
Geometric quantization of 2+1 dimensional pure Yang-Mills theory is studied with focusing on finite large scales. It is previously shown that (Yildirim, 2015, Int. J. Mod. Phys A, 30(7), 1550034), topologically massive Yang-Mills theory exhibits a Chern-Simons splitting behavior at large scales, similar to the topologically massive AdS gravity model in its near Chern-Simons limit. This splitting occurs as two Chern-Simons parts with levels $+k/2$ each, where $k$ is the original Chern-Simons level in the Lagrangian. In the pure Chern-Simons limit, split parts combine to give the original Chern-Simons level. The opposite limit of the gravitational analogue gives an Einstein-Hilbert term with a negative cosmological constant, which can be written as a two half Chern-Simons terms with opposite signs. With this motivation, the gauge theory analogue of this limit is investigated, showing that at finite large distances pure Yang-Mills theory acts like a topological theory that consists of split Chern-Simons parts with levels $+k/2$ and $-k/2$. At very large distances these split terms cancel, making the Yang-Mills theory trivial, as expected, due to the existence of a mass gap. Gauge invariance of the split parts is discussed. It is also shown that this splitting behavior can be exploited to incorporate link invariants of knot theory, just like in the topologically massive case.

\end{abstract}

\begin{center}
\noindent\line(1,0){415}
\end{center}

%-------------------------------------------------------------------------------------------------------------------------
%-------------------------------------------------------------------------------------------------------------------------
\section{Introduction}\label{sec:intro}

It is known that topologically massive AdS gravity(TMAdSG) action can be decomposed into two Chern-Simons(CS) terms\cite{achucarro1986chern, witten1988, Carlip2008272, Carlip2}. For a dynamical metric $\gamma_{\mu\nu}$, the action is\cite{Deser1982372, Deser2}
\al{
\label{eq:tmgS}
S=\int d^3x \left[ -\sqrt{-\gamma}(R-2\Lambda)+\frac{1}{2\mu}\epsilon^{\mu\nu\rho}\left(  \Gamma^\alpha_{\mu\beta}\partial_\nu \Gamma^\beta_{\rho\alpha} +\frac{2}{3}\Gamma^\alpha_{\mu\gamma} \Gamma^\gamma_{\nu\beta} \Gamma^\beta_{\rho\alpha}    \right)  \right].
}
This model is analogous to topologically massive Yang-Mills(TMYM) theory, which is given by the action\footnote{The dimensionless constant factor $k/4\pi$ in front of the Yang-Mills term is not required but inserted for future convenience.}
\al{
\bs
S_{TMYM}=&S_{CS}+S_{YM}\\
=&-\frac{k}{4\pi}\int \limits_{\Sigma \times[t_i,t_f]} {d^3x }\ \epsilon^{\mu\nu\alpha}\ Tr \pa{A_\mu \partial_{\nu} A_{\alpha} + \frac{2}{3}A_\mu A_\nu A_\alpha}\\
&-\frac{k}{4\pi}\frac{1}{4m}\int \limits_{\Sigma\times[t_i,t_f]} {d^3x }\ Tr\  F_{\mu\nu}F^{\mu\nu},
\es
}
where $\Sigma$ is an orientable two dimensional surface, $k$ is the level of CS theory and $m$ is called the topological mass.
\\

With defining
\al{
A^{\pm}{}_{\mu}{}^a{}_b[e]= \omega_{\mu}{}^a{}_b[e] \pm \epsilon^a{}_{bc} e_\mu{}^c,
}
where $e_\mu{}^a$ is the dreibein and $\omega_{\mu}{}^a{}_b[e]$ is the torsion-free spin connection, the action \eqref{eq:tmgS} splits into two CS parts as
\al{
\label{eq:tmgcssplitting}
S[e]=-\frac{1}{2}\pa{1-\frac{1}{\mu}}S_{CS}\big[A^+[e]\big]+\frac{1}{2}\pa{1+\frac{1}{\mu}}S_{CS}\big[A^-[e]\big]
}
where 
\al{
S_{CS}[A]=\frac{1}{2}\int \epsilon^{\mu\nu\rho}\pa{A_\mu{}^a{}_b \partial_\nu A_\rho{}^b{}_a + \frac{2}{3} A_\mu{}^a{}_c A_\nu{}^c{}_b A_\rho{}^b{}_a}.
}
In the \emph{near} CS limit, which is given by small values of $\mu$, the action becomes a sum of two half CS terms as
\al{
\label{eq:tmymlimit}
S[e]\approx\frac{1}{2\mu}S_{CS}\big[A^+[e]\big]+\frac{1}{2\mu}S_{CS}\big[A^-[e]\big].
}
This is analogous to what was observed in ref. \citenum{yildirim} for TMYM theory at large distances. For the following reason, this splitting analogy may seem counter-intuitive. On the gauge theory side of the analogy, the Yang-Mills(YM) term brings scale dependence due to its mass gap, but the gravitational analogue has no such behavior; it can be written purely as a topological theory that consists of CS terms. Scale dependence of the YM term is in the form of exponential decay. Thus, at very large distances, the theory has a very weak scale dependence and can be called almost topological, which can be seen as follows. In ref. \citenum{yildirim}, using methods of Kim, Karabali and Nair\cite{Bos:1989kn, Karabali1996135, Karabali:1996iu, Karabali:1998yq, Karabali:1999ef, Nair:2005iw}, it is shown that the inner product for vacuum states of TMYM theory can be written as
\al{
\label{eq:tmyminnerproduct}
\langle \psi|\psi\rangle_{TMYM_k}=\langle\psi|\psi\rangle_{CS_{k/2}}\langle\psi|\psi\rangle_{CS_{k/2}} +\mc{O}(1/m^2),
}
where labels $k$ and $k/2$ indicate Chern-Simons level. It is known that in TMYM theory, the mass gap is proportional to the topological mass\cite{Karabali:1999ef} $m$ and the distance scale is given in comparison to $k/m$. However, we will continue with fixed $k$ and define the distance scale by $1/m$ throughout the paper. The near CS limit of TMYM theory is given by large values of $m$, as opposed to small values of $\mu$ in its gravitational analogue. It can be seen from \eqref{eq:tmyminnerproduct} that the splitting occurs at large distances compared to $1/m$, by keeping the first order and neglecting all higher order terms in the $1/m$ expansion. An approximation like this is not necessary for the gravitational analogue in order to obtain a topological theory, since it is already topological irrespective of the value of $\mu$. For our interest, the most important difference between TMYM theory and TMAdSG is that the former has a mass gap, which is the reason why topological behavior is observed only at a particular distance scale where $1/m^2$ and higher order terms can be ignored.
\\

Another important limit of the TMAdSG model is $\mu\rightarrow\infty$, which leads to a pure Einstein-Hilbert term with a cosmological constant. In this limit, \eqref{eq:tmgcssplitting} reduces to
\al{
\label{eq:YMlimit}
S[e]=\frac{1}{2}S_{CS}\big[A^-[e]\big]-\frac{1}{2}S_{CS}\big[A^+[e]\big].
}
The gauge theory analogue of the Einstein-Hilbert limit is a pure YM term, which corresponds to small values of $m$ in the TMYM Lagrangian. The goal of this paper is to investigate if a similar CS splitting behavior can be observed in YM theory at large distances. However, one cannot focus on large scales and study pure YM limit of TMYM theory at the same time, since they are given by different limits. For this reason, we will not study YM theory as a small topological mass limit of TMYM theory. Instead, we will start with a pure YM action. If our prediction motivated by the TMAdSG analogy is correct, the inner product of YM theory should be in the form
\al{
\langle \psi|\psi\rangle_{YM}=\langle\psi|\psi\rangle_{CS_{1/2}}\langle\psi|\psi\rangle_{CS_{-1/2}} +\mc{O}(1/m^2).
}

Before testing this prediction by geometrically quantizing pure YM theory, we will quickly go over geometric quantization of CS and TMYM theories to review the methods we are going to use.
%--------------------------------------------------------------------------------------------------------------------------
%--------------------------------------------------------------------------------------------------------------------------

\section{Geometric Quantization of CS Theory}\label{sec:CS}

This section is a review of geometric quantization of non-Abelian CS theory, done by Bos and Nair.\cite{Bos:1989kn,Nair:2005iw}.
\\

The CS action is given by
\al{
S_{CS}=-\frac{k}{4\pi} \int \limits_{\Sigma\times[t_i,t_f]} {d^3x}\ \epsilon^{\mu\nu\alpha}\ Tr \pa{A_\mu \partial_\nu A_\alpha + \frac{2}{3}A_\mu A_\nu A_\alpha}.
}
Here, $A_{\mu}=-iA^a_{\mu}t^a$ where $t^a$ are matrix representatives of the Lie algebra $[t^a,t^b]=if^{abc}t^c$. In the fundamental representation $Tr(t^at^b)=\frac{1}{2}\delta^{ab}$.
\\

CS theory is classically not gauge invariant but at the quantum level it can be made so by restricting $k$ to be an integer. The reasoning is as follows. With a gauge group element $g$, CS action gauge transforms as
\begin{align}
S_{CS}(A)\rightarrow S_{CS}(A^g)=S_{CS}(A)+2\pi k\omega(g),
\end{align}
where $\omega(g)$ is an integer called the winding number. Thus for integer values of $k$, CS path integral does not change, which means it is gauge invariant at the quantum level. 
\\

The field equations for CS are $F_{\mu\nu}=\partial_\mu A_\nu-\partial_\nu A_\mu+[A_\mu,A_\nu] =0$. In the temporal gauge $A_0=0$ with $A_z=\frac{1}{2}(A_1+ iA_2)$ and $A_{\bar{z}}=\frac{1}{2}(A_1- iA_2)$, the momenta are given by
\al{
\Pi^{a z}=\frac{ik}{4\pi} A^a_{\bar{z}}~\ \text{and}~\ \Pi^{a \bar{z}}=-\frac{ik}{4\pi}  A^a_z.
}
The conjugate momenta being given by the gauge fields is a signature behavior of CS theory. 
This leads to an interesting phase space geometry, given by the symplectic two-form
\al{
\label{eq:omegacs}
\Omega=\frac{ik}{2\pi}\int \limits_{\Sigma}\delta A^a_{\bar{z}}\delta A^a_z.
}
It can be seen in \eqr{eq:omegacs} that all phase space coordinates are gauge fields. Later, this property will play a key role in our calculations. 

%-------------------------------------------------------------------------------------------------------------------------

\subsection{Parametrization of Gauge Fields}

When $\Sigma$ is simply connected, we can define a new gauge field $\mc{A}_i$, such that it satisfies \cite{Nair:2005iw}
\al{ 
\label{eq:flatness}
\partial_z A_{\bar{z}}-\partial_{\bar{z}}\mc{A}_z+[\mc{A}_z,A_{\bar{z}}]=0.
} 
With this flatness condition, we can define an $SL(N,\mathbb{C})$ matrix $U$ as
\al{
\label{eq:U}
U(x,0,C)=\mc{P}\ exp\pa{-\underset{C}{\ \ \int_0^x}(A_{\bar{z}}d\bar{z}+\mc{A}_zdz)}.
}
The reason for using the flatness condition \eqref{eq:flatness} in defining $\mc{A}_i$ is to make $U$ invariant under small deformations of the path $C$\cite{Nair:2005iw}. $U$ gauge transforms as $U^g=gU$, where $g\in SU(N)$. From \eqref{eq:U}, it follows that
\al{
\label{eq:par}
A_{\bar z}=-\partial_{\bar z}UU^{-1}~\text{and}~A_z=U^{\dagger-1}\partial_z U^{\dagger}
}
and
\al{
\label{eq:scriptA}
\mc{A}_z=-\partial_z U U^{-1} ~\text{and}~ \mc{A}_{\bar{z}}= U^{\dagger-1}\partial_{\bar{z}} U^{\dagger}.
}

This parametrization will allow us to write the wave-functional in terms of WZW action.

%-------------------------------------------------------------------------------------------------------------------------

\subsection{The Wave-Functional}

At least for the 3D gauge theories we are interested in, the wave-functional can be factorized as $\psi=\phi\chi$. Here, $\phi$ is the scale independent part that satisfies the Gauss' law to ensure that the wave-functional gauge transforms properly, and $\chi$ is the scale dependent part that satisfies the Schr\"odinger's equation. $\phi$ is usually given by some Wess-Zumino-Witten(WZW) action. In the temporal gauge, CS theory Hamiltonian is equal to zero. For this reason, the Schr\"odinger's equation is trivially satisfied and $\chi=1$ is a sufficient solution. This makes the solution scale independent, which is expected since the theory is topological. Thus, the wave-function is given only by the Gauss' law.
\\

To find the wave-functional, we first need to choose a polarization. We choose the holomorphic polarization that relates the pre-quantum ($\Phi$) and quantum ($\psi$) wave-functionals as 
\al{
\Phi[A_z,A_{\bar{z}}]=e^{-\frac{1}{2}K}\psi[A_{\bar{z}}],
}
where $K=\frac{k}{2\pi}\int_{\Sigma} A^a_{\bar{z}}A^a_z$ is the K\"ahler potential for CS theory. In K\"ahler spaces, the pre-quantum inner product can be retained at the quantum level\cite{Nair:2005iw,hall2013} as,
\al{
\label{eq:innerpr}
\langle 1|2 \rangle =\int d\mu(\mc{M})\Phi_1^*\Phi_2 ~\rightarrow~ \int d\mu(\mc{M})e^{-K}\psi_1^*\psi_2
}
where $d\mu(\mc{M)}$ is the Liouville measure given by the symplectic two-form of the theory.
\\

With the polarization we chose, $A_z$ becomes a derivative operator as
\al{
\label{eq:delta}
A^a_z\psi[A^a_{\bar{z}}]=\frac{2\pi}{k} \frac{\delta}{\delta A^a_{\bar{z}}} \psi[A^a_{\bar{z}}].
}

To find $\phi$, we make an infinitesimal gauge transformation on the wave-functional $\psi$ with parameter $\epsilon$, then force the Gauss' law $F_{z\bar{z}}\psi[A_{\bar{z}}]=0$. The wave-functional transforms as
\al{
\bs
\delta_\epsilon \psi[A_{\bar{z}}]=&\int d^2z\ \delta_\epsilon A^a_{\bar{z}}\ \frac{\delta\psi}{\delta A^a_{\bar{z}}} \\
=& \int d^2z\ \epsilon^a \pa{ \partial_{\bar{z}} \frac{\delta}{\delta A^a_{\bar{z}}} + i f^{abc} A^b_{\bar{z}}  \frac{\delta}{\delta A^c_{\bar{z}}} }\psi\\
=& -\frac{k}{2\pi} \int d^2z\ \epsilon^a (F^a_{z\bar{z}} - \partial_z A^a_{\bar{z}}) \psi.
\es
}
Then applying the constraint $F_{z\bar{z}}\psi[A_{\bar{z}}]=0$ leads to
\al{
\label{eq:infg}
\delta_\epsilon \psi[A_{\bar{z}}]= \frac{k}{2\pi} \int d^2z\ \epsilon^a  \partial_z A^a_{\bar{z}} \psi[A_{\bar{z}}].
}
The solution for this condition is given by the WZW action \cite{Polyakov1983121,Polyakov1984223}
\al{
\label{eq:cswf}
\psi[A_{\bar{z}}]=exp\big( kS_{WZW}(U)\big)
}
where
\al{
\bs
S_{WZW}(U)=&\frac{1}{2\pi}\int \limits_{\Sigma}d^2z\ Tr\ \partial_zU \partial_{\bar{z}}U^{-1}\\
&-\frac{i}{12\pi}\int \limits_V d^3x\ \epsilon^{\mu\nu\sigma}\ Tr\ U^{-1}\partial_{\mu}UU^{-1}\partial_{\nu}UU^{-1}\partial_{\sigma}U.
\es
}
Now, we know the wave-functional but we still need to calculate the measure in order to write a complete inner product.
%-------------------------------------------------------------------------------------------------------------------------

\subsection{The Measure}\label{ssec:csmeasure}

The metric on the space of gauge potentials $\mathscr{A}$ is\cite{Karabali1996135}
\al{
\bs
ds^2_{\mathscr{A}}=&\int d^2x\ \delta A^a_i \delta A^a_i=-8\int Tr(\delta A_{\bar{z}} \delta A_z)\\
=& 8 \int Tr[D_{\bar{z}}(\delta U U^{-1})D_z(U^{\dagger -1}\delta U^{\dagger})],
\es
}
where $D_\mu \bullet=\partial_\mu \bullet + [A_\mu,\bullet]$. This metric is related to the Cartan-Killing metric for $SL(N,\mathbb{C})$, which is given by
\al{
ds^2_{SL(N,\mathbb{C})}=8\int Tr[(\delta U U^{-1})(U^{\dagger -1}\delta U^{\dagger})].
}
The volume of  $\mathscr{A}$ is related to the $SL(N,\mathbb{C})$ volume as
\al{
d\mu(\mathscr{A})=det(D_{\bar{z}}D_z)d\mu(U,U^{\dagger}),
}
where $d\mu(U,U^{\dagger})$ is the volume of the space of $SL(N,\mathbb{C})$ matrices.
To make this measure gauge invariant, we define a new $SU(N)$ gauge invariant matrix $H=U^{\dagger}U$; therefore $H \in SL(N,\mathbb{C})/SU(N)$. Using this matrix, the measure becomes
\al{
\label{eq:CSmeasure}
d\mu(\mathscr{A})=det(D_{\bar{z}}D_z)d\mu(H).
}
The determinant is given by the WZW action\cite{Gawedzki:1988hq, Gawedzki:1988nj, Bos:1989kn,Nair:2005iw} as
\al{
det(D_{\bar{z}}D_z)=constant \times e^{2c_AS_{WZW}(H)}
}
where $c_A$ is the quadratic Casimir in the adjoint representation given by the structure constants as $c_A\delta^{ab}=f^{amn}f^{bmn}$.

%--------------------------------------------------------------------------------------------------------------------------

\subsection{The Inner Product}\label{ssec:csinnerproduct}

Using the measure and the wave functional we derived in previous subsections, now we can write the inner product. For K\"ahler polarizations, the inner product is given by
\al{
\langle \psi_1|\psi_2 \rangle=\int d\mu(\mathscr{A})\ e^{-K}\ \psi^*_1\psi^{ }_2.
}
Using the Polyakov-Wiegmann(PW)\cite{Polyakov1983121,Polyakov1984223} identity we get,
\al{
e^{-K}\psi^*\psi=e^{kS_{WZW}(H)}.
}
Now we can finally write the inner product for CS theory as
\al{
\label{eq:csinnerproduct}
\langle \psi|\psi \rangle=\int d\mu(H)\ e^{(2c_A+k)S_{WZW}(H)}.
}
Note that this solution holds only for a simply connected $\Sigma$. For non-simply connected spaces, the parametrization \eqref{eq:par} needs to be modified. For example, for $\Sigma = S^1 \times S^1$, the correct parametrization is
\al{
A_{\bar{z}}=-\partial_{\bar{z}}UU^{-1}+Ui\pi(Im\ \tau)^{-1}aU^{-1},
}
where $a$ is a constant gauge field and $\tau$ is the modular parameter of the torus. A new matrix can be defined as $V=U\ exp[i\pi(Im\ \tau)^{-1}(z-\bar{z})a]$. This allows us to absorb the second term in $V$ and parametrize the gauge field in the same form as before, as $A_{\bar{z}}=-\partial_{\bar{z}}VV^{-1}.$ This modifies the solution to \eqref{eq:infg} as
\al{
\label{eq:toruspsi}
\psi[A_{\bar{z}}]=exp\big( kS_{WZW}(V)\big) \Upsilon(a).
}
Finding $\Upsilon(a)$ is difficult and we will not review its calculation here. The result can be found in ref. \citenum{Bos:1989kn}. 

%--------------------------------------------------------------------------------------------------------------------------
%--------------------------------------------------------------------------------------------------------------------------

\section{Topologically Massive Yang-Mills Theory}\label{sec:TMYM}

This section is a review of K\"ahler quantization of TMYM theory following ref. \citenum{yildirim}.
\\

TMYM theory is simply a mixture of CS and YM theories, as
\al{
\label{eq:actiontmym}
\bs
S_{TMYM}=&S_{CS}+S_{YM}\\
=&-\frac{k}{4\pi}\int \limits_{\Sigma \times[t_i,t_f]} {d^3x }\ \epsilon^{\mu\nu\alpha}\ Tr \pa{A_\mu \partial_{\nu} A_{\alpha} + \frac{2}{3}A_\mu A_\nu A_\alpha}\\
&-\frac{k}{4\pi}\frac{1}{4m}\int \limits_{\Sigma\times[t_i,t_f]} {d^3x }\ Tr\  F_{\mu\nu}F^{\mu\nu}.
\es
} 
Here, $m$ is in mass dimension and called the topological mass. The arbitrary factor $\frac{k}{4\pi}$ is inserted so that pure CS or pure YM limits are given by the value of $m$ alone, independent of $k$. The equations of motion are given by
\al{
\epsilon^{\mu\alpha\beta}F_{\alpha\beta}+\frac{1}{m} D_{\nu}F^{\mu\nu}=0.
}

We would like to introduce suitable phase space coordinates that makes the CS-like behavior more evident. To do this, we define
\al{
\label{eq:tilde0}
\tilde A_{\mu}=A_{\mu}+\frac{1}{2m}\epsilon_{\mu\alpha\beta}F^{\alpha\beta}.
}
$\tilde A_\mu$ transforms like a gauge field since $F^{\alpha\beta}$ is gauge covariant. 
In complex spacial coordinates and in temporal gauge, the conjugate momenta of TMYM theory are 
\al{
\Pi^{a z}=\frac{ik}{4\pi} \tilde A^a_{\bar{z}}~\ \text{and}~\ \Pi^{a \bar{z}}=-\frac{ik}{4\pi}  \tilde A^a_z,
}
with
\al{
\label{eq:Az}
\tilde A_{z}=A_{z}+E_{z}~\ \text{and}~\ \tilde A_{\bar{z}}=A_{\bar{z}}+E_{\bar{z}},
}
where
\al{
\label{eq:E}
E_{z}=\frac{i}{2m}F^{0\bar{z}} \ ~\text{and}~\  E_{\bar{z}}=-\frac{i}{2m}F^{0z}.
}
Dropping the tilde sign is the equivalent of taking the pure CS limit, by killing the $E$-field contributions to momenta from the YM term.

%---------------------------------------------------------------------------------------------------------------------- 

\subsection{Phase Space Geometry}

With the new coordinates we defined, the symplectic two-form for this theory is
\al{
\label{eq:omegatmym}
\Omega=&\frac{ik}{4\pi}\int \limits_{\Sigma}(\delta \tilde A^a_{\bar{z}} \delta A^a_z+\delta A^a_{\bar{z}}\delta \tilde A^a_z ).
}
Comparing this symplectic two-form with \eqr{eq:omegacs} reveals that TMYM theory has a \emph{CS-like} phase space geometry that consists of two half-CS parts. Here, the words ``\emph{CS-like}'' are used in the sense that all phase space coordinates transform like gauge fields. This behavior is even more evident in the following phase space coordinates:
\al{
B_z=\frac{1}{2}(A_1+i\tilde A_2)\ ~\text{and}~\ C_z=\frac{1}{2}(\tilde A_1+i A_2).
}
In these coordinates, the symplectic two-form of TMYM theory can be written as
\al{
\label{eq:omegaBCtmym}
\Omega=\frac{ik}{4\pi}\int \limits_{\Sigma}(\delta B^a_{\bar{z}} \delta B^a_z+\delta C^a_{\bar{z}}\delta C^a_z).
}
Here, both $B_i$ and $C_i$ transform like gauge fields. It can be seen that there are separate CS sectors associated with $B$ and $C$ gauge fields. Pure CS limit is given by $m\rightarrow\infty$ or equivalently $B_i=C_i$, which reduces \eqref{eq:omegaBCtmym} to \eqref{eq:omegacs}.
\\

One might ask if it is possible to write any gauge theory phase space in CS-like pieces. The answer is obviously no for scalar field theories or gauge theories in different number of space-time dimensions. But CS is such an essential 3D gauge theory, it can be used as a geometric setup for other 3D gauge theories. This was shown classically in ref. \citenum{Lemes:1998md} by Lemes et.al. They showed that it is possible to write 3D gauge theories as some form of CS theory via local redefinitions of gauge fields. At quantum level, different factors come into play, such as existence of a mass gap, making these classical equivalences generally invalid. Despite the differences at quantum level, CS phase space appears to be the fundamental building block of phase spaces of other 3D gauge theories. 

%-------------------------------------------------------------------------------------------------------------------------

\subsection{The Wave-Functional}

In this subsection, we will apply the methods outlined in \autoref{sec:CS} on TMYM theory. Once again, we choose the K\"ahler polarization. In this polarization, the pre-quantum and quantum wave-functionals are related by 
\al{
\Phi[A_z,A_{\bar{z}},\tilde A_z, \tilde A_{\bar{z}}]=e^{-\frac{1}{2}K}\psi[A_{\bar{z}},\tilde A_{\bar{z}}],
} 
where $K=\frac{k}{4\pi}\int_\Sigma (\tilde A^a_{\bar{z}} A^a_z+ A^a_{\bar{z}} \tilde A^a_z )$ is the K\"ahler potential. The quantum operators are given by
\al{
\label{eq:deltatmym}
A^a_z\psi=\frac{4\pi}{k} \frac{\delta}{\delta \tilde A^a_{\bar{z}}} \psi\ ~\text{and}\ ~\tilde A^a_z\psi=\frac{4\pi}{k} \frac{\delta}{\delta A^a_{\bar{z}}}\psi .
}
Because of the $ \tilde A_{\bar{z}}$ dependence of $\psi$, an infinitesimal gauge transformation of it has an additional term compared to the CS case, as 
\al{
\label{eq:infgauge}
\delta_\epsilon \psi[A_{\bar{z}},\tilde{A}_{\bar{z}}]=\int d^2z\ \pa{ \delta_\epsilon A^a_{\bar{z}} \frac{\delta\psi}{\delta A^a_{\bar{z}}}  +\ \delta_\epsilon \tilde{A}^a_{\bar{z}} \frac{\delta\psi}{\delta \tilde{A}^a_{\bar{z}}}}.
}
Now using \eqr{eq:deltatmym} and $\delta A^a_{\bar{z}}=D_{\bar{z}}\epsilon^a$, $\delta \tilde{A}^a_{\bar{z}}=\tilde{D}_{\bar{z}}\epsilon^a$, where $\tilde D_\mu \bullet=\partial_\mu \bullet + [\tilde A_\mu,\bullet]$, the transformation takes the form:
\al{
\bs
\delta_\epsilon \psi=& \int d^2z\ \epsilon^a \pa{\tilde{D}_{\bar{z}}\frac{\delta}{\delta \tilde{A}^a_{\bar{z}}} + D_{\bar{z}}\frac{\delta}{\delta A^a_{\bar{z}}} } \psi\\
=&\frac{k}{4\pi} \int d^2z\ \epsilon^a \pa { \partial_z \tilde{A}^a_{\bar{z}}+ \partial_z A^a_{\bar{z}} -2F_{z\bar{z}}- D_z E_{\bar{z}}+D_{\bar{z}}E_z  }\psi.
\es
}
The last three terms cancel via Gauss' law $(2F_{z\bar{z}}+D_z E_{\bar{z}}-D_{\bar{z}}E_z)\psi=0$, leading to 
\al{
\label{eq:infg2}
\delta_\epsilon \psi= \frac{k}{4\pi} \int d^2z\ \epsilon^a \pa{\partial_z \tilde A^a_{\bar{z}}+\partial_z A^a_{\bar{z}} } \psi.
}
Similar to \eqr{eq:infg}, \eqr{eq:infg2} is solved by the WZW action. To do this, we need to parametrize $\tilde A$ similar to $A$, using a new SL(N,$\mathbb{C}$) matrix
\al{
\label{eq:Utilde}
\tilde U(x,0,C)=\mc{P}\ exp\pa{-\underset{C}{\ \ \int_0^x}(\tilde A_{\bar{z}}d\bar{z}+ \tilde{\mc{A}}_zdz)}.
}
Thus, $\tilde A_{\bar z}=-\partial_{\bar z}\tilde U \tilde U^{-1}$, $\tilde A_z=\tilde U^{{\dagger}{-1}}\partial_z \tilde U^{\dagger}$ and $\tilde{\mc{A}}_{z}=-\partial_{z}\tilde U \tilde U^{-1}$, $\tilde{\mc{A}}_{\bar z}=\tilde U^{{\dagger}{-1}}\partial_{\bar z} \tilde U^{\dagger}$. Now with $\tilde U$, the wave-functional is given by
\al{
\label{eq:wftmym}
\psi[A_{\bar{z}},\tilde{A}_{\bar{z}}]=\phi\chi,\ \ \text{with}\ \ \phi=\exp\pb{\frac{k}{2}\big(S_{WZW}(\tilde U)+S_{WZW}( U)\big)}.
}
Unlike CS theory, the scale dependent factor $\chi$ is not equal to 1, since TMYM theory not topological. $\chi$ can be found by solving the Schr\"odinger's equation. 
\\

In the temporal gauge, there is no CS contribution to the TMYM Hamiltonian, thus, it is equal to the pure YM Hamiltonian. The $E$-fields satisfy $[E^a_z(x),E^b_{\bar{z}}(x')]=-\frac{8\pi}{k} \delta^{ab}\delta^{(2)}(x-x')$. Therefore $E$-fields act like creation and annihilation operators and the Hamiltonian must be normal ordered\cite{Grignani1997360}. The normal ordered Hamiltonian in Euclidean geometry is given by
\al{
\mc{H}=\frac{mk}{4\pi}E^a_{\bar{z}} E^a_z +\frac{4\pi}{mk} B^a B^a,
}
where $B=\frac{ik}{2\pi}F^{z\bar{z}}$.
\\

The vacuum wave-functional is given by $\mc{H}\psi=0$. However, this equation is highly non-linear and not solvable without any approximation. Fortunately, we are only interested in large distance behavior of the theory, which corresponds to low energies. Neglecting the potential energy term is standard practice in these type of cases\cite{Karabali:1999ef, Grignani1997360}. The reasoning is as follows. Since the energy is low, charges move very slowly and do not create significant magnetic fields\cite{Grignani1997360}. With this approximation, now the vacuum wave-functional is given by $E^a_z\psi=0$.
\\

Before we tackle the Schr\"odinger's equation, we should note that derivatives of $\phi$ are given by the fields $\mc{A}$ and $\mc{\tilde{A}}$ that were defined in \eqref{eq:U} and \eqref{eq:Utilde}, as
\al{
\label{eq:phi}
\tilde A^a_z\phi=\frac{4\pi}{k}\frac{\delta \phi}{\delta A^a_{\bar{z}}}=\mc{A}^a_z \phi\  ~\text{and}~\  A^a_z\phi=\frac{4\pi}{k}\frac{\delta \phi}{\delta \tilde{A}^a_{\bar{z}}}=\tilde{\mc{A}}^a_z \phi.
}
We can define a gauge covariant field $\mc{E}$, analogous to $E$, that satisfies $\mc{\tilde{A}}=\mc{A}+\mc{E}$. By using \eqref{eq:phi}, we obtain
\al{
\label{eq:EE}
E^a_z\psi=-\mc{E}^a_z\psi+\frac{4\pi}{k}\pa{\frac{\delta \chi}{\delta A^a_{\bar{z}}}-\frac{\delta \chi}{\delta \tilde{A}^a_{\bar{z}}}}\phi.
}
Now, the solution to $E^a_z\psi=0$ is given by
\al{
\label{eq:chi0}
\bs
\chi=&exp \pa{-\frac{k}{8\pi}\int \limits_{\Sigma} (\tilde{A}^a_{\bar{z}}-A^a_{\bar{z}})\mc{E}^a_z}=exp \pa{-\frac{k}{8\pi}\int \limits_{\Sigma}E^a_{\bar{z}}\mc{E}^a_z}.
\es
}
Both $E^a_{\bar{z}}$ and $\mc{E}^a_z$ are in the order of $1/m$, which leads to
\al{
\chi^*\chi=1+\mc{O}(1/m^2).
}
This means that at large distances where $1/m^2$ and higher order terms can be neglected, $\chi^*\chi$ can taken to be unity.

%------------------------------------------------------------------------------------------------------------------------

\subsection{The Measure}

Similar to the analysis in \autoref{ssec:csmeasure}, we will start with the metric of the space of gauge fields, which is given by the symplectic two-form \eqref{eq:omegatmym}, as
\al{
\bs
ds^2_{\mathscr{A}}=&-4\int Tr(\delta \tilde{A}_{\bar{z}} \delta A_z+\delta A_{\bar{z}}\delta \tilde{A}_z)\\
=&\ 4 \int Tr[\tilde{D}_{\bar{z}}(\delta \tilde{U} \tilde{U}^{-1})D_z(U^{\dagger -1}\delta U^{\dagger})+D_{\bar{z}}(\delta U U^{-1})\tilde{D}_z(\tilde{U}^{\dagger -1}\delta \tilde{U}^{\dagger})].
\es
}
For this metric, the gauge invariant measure is
\al{
\label{eq:meas}
d\mu(\mathscr{A})=det(\tilde{D}_{\bar{z}}D_z) det(D_{\bar{z}}\tilde{D}_z)d\mu(\tilde{U}^{\dagger}U)d\mu(U^{\dagger}\tilde{U}).
}
Once again, the determinants are given by WZW action as
\al{
det(\tilde{D}_{\bar{z}}D_z) det(D_{\bar{z}}\tilde{D}_z)=constant \times e^{2c_A\big(S_{WZW}(\tilde{U}^{\dagger}U)+S_{WZW}(U^{\dagger}\tilde{U})\big)}.
}
We can define another $SU(N)$ gauge invariant matrix $N=\tilde{U}^{\dagger}U$ to simplify the notation. Now the measure becomes
\al{
\label{eq:measure}
d\mu(\mathscr{A})=constant \times e^{2c_A\big(S_{WZW}(N)+S_{WZW}(N^{\dagger})\big)} d\mu(N)d\mu(N^{\dagger}).
}

%------------------------------------------------------------------------------------------------------------------------

\subsection{The Inner Product}

The inner product for K\"ahler polarizations is given by \eqr{eq:innerpr}. Using PW identity, the integrand can be written as
\al{
e^{-K_{TMYM}}\psi_{TMYM}^*\psi^{ }_{TMYM}=e^{\frac{k}{2}\big(S_{WZW}(N)+S_{WZW}(N^{\dagger})\big)}\chi^*\chi.
}

As we have said before, $\chi^*\chi\approx1$ at large scales compared to $1/m$. This leads to an almost topological theory that consists of two half-CS parts as
\al{
\bs
\label{eq:tmyminnerproduct2}
\langle \psi|\psi\rangle_{TMYM_k}{\approx}& \int d\mu(N)d\mu(N^{\dagger})\ e^{(2c_A+\frac{k}{2})\big(S_{WZW}(N)+S_{WZW}(N^{\dagger})\big)}\\
=&\langle\psi|\psi\rangle_{CS_{k/2}}\langle\psi|\psi\rangle_{CS_{k/2}}.
\es
} 

%---------------------------------------------------------------------------------------------------------------------- 
 
\subsection{The $ \Sigma = S^1 \times S^1 $ Case} 

CS splitting behavior seems to be independent of $\Sigma$ and work for non-simply connected cases, such as $\Sigma=S^1\times S^1$.  In that case, similar to \eqref{eq:toruspsi}, TMYM wave-functional for  $\Sigma=S^1\times S^1$ is
\al{
\label{eq:toruspsi}
\psi[A_{\bar{z}},\tilde{A}_{\bar{z}}]=exp\pb{\frac{k}{2}(S_{WZW}(\tilde V)+S_{WZW}( V))}\Upsilon(\tilde{a})\Upsilon(a)\chi.
}
It can be seen in \eqref{eq:toruspsi} that, at large distances where $\chi \approx 1$, CS splitting can be observed. This is not surprising since it satisfies the gauge transformation condition \eqref{eq:infg2}, which is the doubled version of \eqref{eq:infg}.

%---------------------------------------------------------------------------------------------------------------------- 

\subsection{CS Splitting and Gauge Invariance}\label{sec:gauge1}

Although $\langle \psi|\psi\rangle_{TMYM_k}$ is gauge invariant in its own right for all integer values of $k$, it seems like gauge invariance is broken on the right hand side of \eqref{eq:tmyminnerproduct2} for odd values of $k$. This is not true, since both half-CS terms separately bring an extra $\pi k \omega$ term as they transform, where $\omega$ is the winding number. They add up to $2 \pi k \omega$, which does not change the value of the path integral. This transformation can be written as
\al{
\frac{1}{2}S_{CS}(B)+\frac{1}{2}S_{CS}(C)\rightarrow \frac{1}{2}S_{CS}(B)+\frac{1}{2}S_{CS}(C)+2\pi k \omega.
}
Here, transformation of both CS terms bring the same winding number because both $B$ and $C$ gauge fields transform together with the same gauge group element. If separate gauge invariance of the split CS halves is required for any reason, such as writing TMYM observables in terms of split-CS observables, then $k$ must be forced to take only even values.

%--------------------------------------------------------------------------------------------------------------------------
%--------------------------------------------------------------------------------------------------------------------------

\section{Yang-Mills Theory In 2+1 Dimensions}\label{sec:YM}

The YM action is given by
\al{
\label{eq:actionym}
S_{YM}=-\frac{k}{4\pi}\frac{1}{4m}\int \limits_{\Sigma\times[t_i,t_f]} {d^3x }\ Tr\  (F_{\mu\nu}F^{\mu\nu}).
} 
where $m$ is in mass dimensions. The arbitrary constant $\frac{k}{4\pi}$ is not really necessary but will be helpful in relating some YM observables to CS observables. Here, unlike CS theory, there is no need to introduce a restriction on $k$ to obtain gauge invariance; YM Lagrangian is already gauge invariant. At this point, $k$ is allowed to take non-integer values. 
\\

The equations of motion are given by $D_{\nu}F^{\mu\nu}=0$. In the temporal gauge $A_0=0$ with complex coordinates, the conjugate momenta are
\al{
\Pi^{a z}=\frac{k}{8\pi m}F^{a0z}~\ \text{and}~\ \Pi^{a \bar{z}}= \frac{k}{8\pi m}  F^{a0 \bar{z}}.
}
With $E_{z}=\frac{i}{2m}F^{0\bar{z}}$ and $E_{\bar{z}}=-\frac{i}{2m}F^{0z}$, the momenta can be rewritten as
\al{
\Pi^{a z}=\frac{ik}{4\pi} E^a_{\bar{z}}~\ \text{and}~\ \Pi^{a \bar{z}}=-\frac{ik}{4\pi} E^a_z.
}
With these coordinates, the symplectic two form is given by
\al{
\label{eq:omegaym}
\Omega=\frac{ik}{4\pi}\int \limits_{\Sigma}(\delta E^a_{\bar{z}} \delta A^a_z+\delta A^a_{\bar{z}}\delta E^a_z).
}
The the Gauss' law is given by $(D_z E_{\bar{z}} - D_{\bar{z}}E_z)^a=0$, which is also one of the equations of motion.
\\

To show how the phase space CS splits, we want to write $\Omega$ in two CS-like parts. To do this, we will use coordinates
\al{
\label{eq:tildehatdef}
\tilde A_{\mu}=A_{\mu}+\frac{1}{2m}\epsilon_{\mu\alpha\beta}F^{\alpha\beta}
~\text{and}~
\hat A_{\mu}=A_{\mu}-\frac{1}{2m}\epsilon_{\mu\alpha\beta}F^{\alpha\beta}
}
or equally
\al{
\tilde A_i=A_i+E_i\ ~\text{and}~\ \hat A_i=A_i-E_i.
}
Since $E_i$ is gauge covariant, $\tilde A_i$ and $\hat A_i$ transform like gauge fields.
Now, $\Omega$ can be written only in terms of gauge fields as
\al{
\label{eq:omegaym2}
\Omega=\frac{ik}{4\pi}\int \limits_{\Sigma}(\delta \tilde A^a_{\bar{z}} \delta A^a_z-\delta A^a_{\bar{z}}\delta \hat A^a_z).
}
The phase space is K\"ahler with the potential 
\al{
\label{eq:kahlerpot}
K=\frac{k}{4\pi}\int \limits_{\Sigma}(\tilde A^a_{\bar{z}} A^a_z-A^a_{\bar{z}}\hat A^a_z).
}
Similar to $\tilde D_i$, using $\hat A_i$ as connection, a new covariant derivative  $\hat D_i$ can be defined that is going to be useful later.
\\

Chern-Simons splitting of the phase space is more evident with the coordinates $B_z=\frac{1}{2}(\tilde A_1+iA_2)$ and $C_z=\frac{1}{2}(A_1+i\hat A_2)$. Now, the symplectic two-form can be written as
\al{
\label{eq:omegaBC}
\Omega=\frac{ik}{4\pi}\int \limits_{\Sigma}(\delta B^a_{\bar{z}} \delta B^a_z-\delta C^a_{\bar{z}}\delta C^a_z).
}
Once again, both $B_i$ and $C_i$ transform like non-abelian gauge fields. It can be seen that \eqref{eq:omegaBC} consists of two CS phase spaces with levels $k/2$ and $-k/2$. In the $m\rightarrow \infty$ limit, $B=C$ and this cancellation leads to a trivial phase space.

%------------------------------------------------------------------------------------------------------------------------

\subsection{The Wave-Functional}

Choosing the holomorphic polarization gives $\Phi[A_z, A_{\bar{z}},\hat A_z, \tilde A_{\bar{z}}]=e^{-\frac{1}{2}K}\psi[ A_{\bar{z}},\tilde A_{\bar{z}}]$, where $K$ is the K\"ahler potential given by \eqref{eq:kahlerpot}, $\Phi$ and $\psi$ are the pre-quantum and quantum wave-functionals. This polarization choice is crucial in studying the large distance behavior, since real polarizations cannot resolve large scales because of technical difficulties in solving the Schr\"odinger's equation\cite{yildirim}.
\\

From \eqref{eq:omegaym2}, upon quantization we can write
\al{
\label{eq:deltaym}
A^a_z\psi=\frac{4\pi}{k} \frac{\delta}{\delta \tilde A^a_{\bar{z}}}\psi \ ~\text{and}\ ~\hat A^a_z\psi=-\frac{4\pi}{k} \frac{\delta}{\delta A^a_{\bar{z}}}\psi.
}
Now, we make an infinitesimal gauge transformation on $\psi$,
\al{
\label{eq:infgauge}
\delta_\epsilon \psi[A_{\bar{z}},\tilde{A}_{\bar{z}}]=\int d^2z\ \pa{ \delta_\epsilon A^a_{\bar{z}} \frac{\delta\psi}{\delta A^a_{\bar{z}}}  +\ \delta_\epsilon \tilde{A}^a_{\bar{z}} \frac{\delta\psi}{\delta \tilde{A}^a_{\bar{z}}}}.
}
Using \eqr{eq:deltaym}, $\delta A^a_{\bar{z}}=D_{\bar{z}}\epsilon^a$ and $\delta \tilde{A}^a_{\bar{z}}=\tilde{D}_{\bar{z}}\epsilon^a$, \eqr{eq:infgauge} can be rewritten as
\al{
\label{eq:infgauge2}
\bs
\delta_\epsilon \psi=& \int d^2z\ \epsilon^a \pa{ D_{\bar{z}}\frac{\delta}{\delta A^a_{\bar{z}}} + \tilde{D}_{\bar{z}}\frac{\delta}{\delta \tilde{A}^a_{\bar{z}}} } \psi\\
=&\frac{k}{4\pi} \int d^2z\ \epsilon^a \pa { \tilde D_{\bar{z}}A^a_z - D_{\bar{z}}\hat A^a_z }\psi.
\es
}
Using \eqref{eq:tildehatdef}, \eqref{eq:infgauge2} becomes
\al{
\delta_\epsilon \psi=&\frac{k}{4\pi} \int d^2z \epsilon^a \pa { \partial_z E^a_{\bar{z}} - D_z E^a_{\bar{z}} + D_{\bar{z}}E^a_z }\psi.
}
Applying the Gauss' law $(D_z E_{\bar{z}} - D_{\bar{z}}E_z)^a\psi=0$ cancels the last two terms, leaving
\al{
\label{eq:infgauge3}
\bs
\delta_\epsilon \psi=&\frac{k}{4\pi} \int d^2z\ \epsilon^a \pa { \partial_z E^a_{\bar{z}} }\psi\\
=&\frac{k}{4\pi} \int d^2z\ \epsilon^a \pa { \partial_z \tilde A^a_{\bar{z}}-\partial_z A^a_{\bar{z}} }\psi\\
=&\frac{k}{4\pi} \int d^2z\ \epsilon^a \pa {\partial_z  A^a_{\bar{z}}-\partial_z \hat A^a_{\bar{z}} }\psi.
\es
}
This condition is solved by $\psi=\phi\chi$ with
\al{
\label{eq:wfym1}
\phi[A_{\bar{z}},\tilde{A}_{\bar{z}}]=exp\pb{\frac{k}{2}\big(S_{WZW}(\tilde U)-S_{WZW}( U)\big)},
}
or equally
\al{
\label{eq:wfym2}
\phi[A_{\bar{z}},\hat{A}_{\bar{z}}]=exp\pb{\frac{k}{2}\big(S_{WZW}(U)-S_{WZW}( \hat U)\big)}.
}
The equivalence of \eqref{eq:wfym1} and \eqref{eq:wfym2} can be shown by using the Polyakov-Wiegmann(PW) identity\cite{Polyakov1983121,Polyakov1984223} with $\tilde U=UM$ and $\hat U=UM^{-1}$.

Holomorphic components of gauge fields acting on $\phi$ gives $\mc{A}$ and $\tilde{\mc{A}}$ fields, as
\al{
\label{eq:phi2}
A^a_z\phi=\frac{4\pi}{k}\frac{\delta \phi}{\delta \tilde A^a_{\bar{z}}}=\tilde{\mc{A}}^a_z \phi\  ~\text{and}~\  \hat A^a_z\phi=-\frac{4\pi}{k}\frac{\delta \phi}{\delta A^a_{\bar{z}}}=\mc{A}^a_z \phi.
}

%------------------------------------------------------------------------------------------------------------------------

\subsection{The Schr\"odinger's Equation}\label{sec:schrodinger}

YM Hamiltonian is given by
\al{
\mc{H}=T+V\ ~\text{with}~\ T=\frac{m}{\alpha}E^a_{\bar{z}} E^a_z\ ~\text{and}~\  V=\frac{\alpha}{m} B^a B^a,
}
where $\alpha=\frac{4\pi}{k}$, $B=\frac{ik}{2\pi}F^{z\bar{z}}$. We are interested in finding the vacuum wave-functional, which is given by $\mc{H}\psi=0$, or with using Euclidean metric,
\al{
\label{eq:Hamilt}
E^a_{\bar{z}}E^a_z \psi +\frac{1}{64m^2} F^a_{z\bar{z}} F^a_{z\bar{z}} \psi =0.
}
For both YM and TMYM theories, $\chi$ is typically in the form\footnote{Except for very short distances.} of $e^{-\frac{1}{m^2}\int F^2}$\cite{Karabali:1998yq,Karabali:1999ef,yildirim}, since they have the same Hamiltonian in the temporal gauge. This decay behavior originates from the existence of a mass gap and it should be valid here as well. To check this prediction, we will first focus on the potential energy term. Using \eqref{eq:phi2}, the $B$-field acting on $\psi$ is
\al{
\label{eq:B}
\bs
F^a_{z\bar{z}}\psi=&(\partial_z A^a_{\bar{z}}-D_{\bar{z}}A^a_z)\psi\\
=&D_{\bar{z}}(\mc{A}^a_z-A^a_z)\psi\\
=&D_{\bar{z}}\pa{-\mc{E}^a_z\psi-\frac{4\pi}{k}\frac{\delta \chi}{\delta \tilde{A}^a_{\bar{z}}}\phi},
\es
}
where $\mc{E}_z$ is given by $\tilde{\mc{A}}_z-\mc{A}_z$. If we consider only the potential energy term, $F^a_{z\bar{z}}\psi=0$ can be solved by
\al{
\label{eq:chiB}
\chi=exp\pa{-\frac{k}{4\pi}\int \limits_{\Sigma}E^a_{\bar{z}}\mc{E}^a_z}=exp\pa{-\frac{k}{4\pi}\int \limits_{\Sigma}(\tilde A^a_{\bar{z}}-A^a_{\bar{z}})\mc{E}^a_z}=1+\mc{O}(1/m^2).
}
Now, to separately study the kinetic energy term, $E_z\psi$ needs to be calculated. Using \eqref{eq:phi2}, we can write
\al{
\bs
E^a_z\psi=&(A^a_z-\hat A^a_z)\phi\chi\\
=&\mc{E}^a_z \psi + \frac{4\pi}{k}\phi \pa{\frac{\delta\chi}{\delta \tilde A^a_{\bar{z}}}+\frac{\delta\chi}{\delta A^a_{\bar{z}}}}.
\es
}
$\chi$ is a function of both $A_{\bar{z}}$ and $\tilde A_{\bar{z}}$ and it is gauge invariant by definition. The only way to ensure gauge invariance with these variables is to have $\chi[\tilde A^a_{\bar{z}},A^a_{\bar{z}}]=\chi[\tilde A^a_{\bar{z}}-A^a_{\bar{z}}]=\chi[E^a_{\bar{z}}]$, which can be seen clearly in \eqref{eq:chiB}. We can then write
\al{
\frac{\delta\chi}{\delta \tilde A^a_{\bar{z}}}=-\frac{\delta\chi}{\delta A^a_{\bar{z}}}.
}
This leads to 
\al{
E^a_{\bar{z}}E^a_z\psi = E^a_{\bar{z}}\mc{E}^a_z\psi.
}
Thus, as far as only the kinetic energy term is concerned, a constant $\chi$ is sufficient. This result is in agreement with Kim, Karabali and Nair's work\cite{Karabali:1998yq}. If we neglect the potential energy term, the vacuum condition forces $\mc{E}^a_z$ to be zero.  It can be seen from \eqref{eq:chiB} that, $\mc{E}^a_z=0$ is also consistent with neglecting the $B$-field contribution.
\\

Our goal is to study the large distance behavior of the theory by neglecting second and higher order terms in $1/m$. However, we may comment on shorter distances as well. When the potential term is not neglected, we expect the full solution for \eqref{eq:Hamilt} to be an interpolation between the kinetic energy eigenstate and potential energy eigenstate. Since neither of these states has a first order term in $1/m$, even if we do not neglect the potential term, there should not be any first order contribution. This result is consistent with ref. \citenum{Karabali:1998yq}. Thus, for the scale that we focus on, $\chi=1$ is a sufficient solution, just like in TMYM theory\cite{yildirim}. 

%------------------------------------------------------------------------------------------------------------------------

\subsection{The Measure}

From \eqref{eq:kahlerpot}, we can write the metric of the space of gauge fields $\mathscr{A}$ as
\al{
\bs
ds^2_{\mathscr{A}}=&-4\int Tr(\delta \tilde{A}_{\bar{z}} \delta A_z-\delta A_{\bar{z}}\delta \hat{A}_z)\\
=&\ 4 \int Tr[\tilde{D}_{\bar{z}}(\delta \tilde{U} \tilde{U}^{-1})D_z(U^{\dagger -1}\delta U^{\dagger})-D_{\bar{z}}(\delta U U^{-1})\hat{D}_z(\hat{U}^{\dagger -1}\delta \hat{U}^{\dagger})].
\es
}
Similar to the analyses in previous sections, the gauge invariant measure is given by
\al{
\label{eq:meas}
d\mu(\mathscr{A})=det(\tilde{D}_{\bar{z}}D_z) det(D_{\bar{z}}\hat{D}_z)d\mu(\hat{U}^{\dagger}U)d\mu(U^{\dagger}\tilde{U}).
}
The determinants are once again given by WZW actions, as
\al{
det(\tilde{D}_{\bar{z}}D_z) det(D_{\bar{z}}\hat{D}_z)=const. \times e^{2c_A\big(S_{WZW}(U^{\dagger}\tilde{U})+S_{WZW}(\hat{U}^{\dagger}U)\big)}.
}
To simplify the notation, we will continue with $H_1=U^{\dagger}\tilde{U}$ and $H_2=\hat{U}^{\dagger}U$. These two matrices are $SU(N)$ gauge invariant and belong to the coset $SL(N,\mathbb{C})/SU(N)$. Now, the gauge invariant measure can be written as
\al{
\label{eq:meas2}
d\mu(\mathscr{A})=e^{2c_A\big(S_{WZW}(H_1)+S_{WZW}(H_2)\big)}d\mu(H_1)d\mu(H_2).
}

%-------------------------------------------------------------------------------------------------------------------------

\subsection{CS Splitting and Gauge Invariance }

For holomorphic polarization, the inner product is given by \eqr{eq:innerpr}. Using \eqref{eq:kahlerpot}, \eqref{eq:wfym1} and \eqref{eq:wfym2} with PW identity gives
\al{
\label{eq:psipsi}
e^{-K}\ \psi^*\psi^{ }=e^{\frac{k}{2}\big(S_{WZW}(H_1)-S_{WZW}(H_2)\big)}\chi^*\chi^{ }.
}
As we have seen in \autoref{sec:schrodinger},
\al{
\label{eq:chiorder}
\chi^*\chi^{ }=1+\mc{O}(1/m^2).
}
Now, using \eqref{eq:meas2}, \eqref{eq:psipsi} and \eqref{eq:chiorder}, we can write the inner product for the vacuum state as
\al{
\label{eq:YMinnerproduct}
\langle \psi|\psi \rangle=\int d\mu(H_1)d\mu(H_2)e^{\pa{2c_A+\frac{k}{2}}S_{WZW}(H_1)+\pa{2c_A-\frac{k}{2}}S_{WZW}(H_2)}+\mc{O}(1/m^2).
}
The inner product for CS theory is given by \eqr{eq:csinnerproduct}. It can be seen that the YM inner product \eqref{eq:YMinnerproduct} consists of two CS parts with levels $k/2$ and $-k/2$ that cancel as $m\rightarrow\infty$, since $H_1=H_2=H$ in this limit. Using \eqr{eq:csinnerproduct}, YM inner product can be written as
\al{
\label{eq:YMsplit}
\langle \psi|\psi\rangle_{YM_k}=\langle \psi|\psi \rangle_{CS_{k/2}}\langle \psi|\psi \rangle_{CS_{-k/2}}+\mc{O}(1/m^2).
}
It appears that YM inner product splits into two CS terms at large distances, just as we predicted by studying its gravitational analogue \eqref{eq:YMlimit}. 

The gauge invariance issue of the right hand side of \eqref{eq:YMsplit} can be solved similar to our analysis in \autoref{sec:gauge1}. Under a gauge transformation, two split CS terms bring $\pi k \omega$ and $-\pi k \omega$ as
\al{
\frac{1}{2}S_{CS}(B)-\frac{1}{2}S_{CS}(C)\rightarrow \frac{1}{2}S_{CS}(B)-\frac{1}{2}S_{CS}(C)+\pi k \omega-\pi k \omega,
}
which leads to gauge invariance for all values of $k$. Although no restriction for $k$ is required here, as we will see in the next section, writing YM observables at large scales in terms of CS observables can only be done for even values of $k$.

%-------------------------------------------------------------------------------------------------------------------------

\section{Wilson Loops}

In ref. \citenum{yildirim}, CS splitting was utilized to incorporate knot theory. A similar calculation can be done here, for pure YM theory at large distances. 
\\

In the temporal gauge, the Wilson loop operator is
\al{ 
W_R(C)=Tr_R\ \mc{P}\ e^{-\oint \limits_c (A_zdz+A_{\bar{z}}d\bar{z})}.
}
Expanding the path ordered exponential for $W_R(C)\psi$ is complicated, since $A_z$ is a derivative operator. This problem is solved in ref. \citenum{yildirim} by replacing $A_z$ with $\mc{A}_z$ to define a Wilson loop-like observable $\mc{W}_R(C)$ as
\al{
\mc{W}_R(C)=Tr_R\ \mc{P}\ e^{-\oint \limits_c (\mc{A}_zdz+A_{\bar{z}}d\bar{z})}= Tr_R\ U(x,x,C).
}
This replacement makes the operator invariant under small deformations of the path $C$ on $\Sigma$. However, it does not prevent doing deformations in the time direction to obtain skein relations. For pure CS theory, once the Gauss' law constraint is applied, $A_z$ becomes equal to $\mc{A}_z$. Thus, the $A_z \rightarrow \mc{A}_z$ replacement procedure gives the Wilson loop a CS-like behavior irrespective of the Lagrangian. 
\\

Using $\tilde A$, we can define a dual operator
\al{
\mc{T}_R(C)=Tr_R\ \mc{P}\ e^{-\oint \limits_c (\tilde{\mc{A}}_zdz+\tilde{A}_{\bar{z}}d\bar{z})}= Tr_R\ \tilde U(x,x,C).
}
After the derivative operators are replaced with $\mc{A}$ and $\tilde{\mc{A}}$, the new loop operators $\mc{W}_R(C_1)$ and $\mc{T}_R(C_2)$ commute with each other, as well as with $\psi[U,\tilde U]$. Here, $C_1$ and $C_2$ are non-intersecting closed curves.
\\

Since $\mc{W}_R(C_1)$ and $\mc{T}_R(C_2)$ are gauge invariant and the theory is given by the WZW action, these loop operators can be written in terms of gauge invariant WZW currents via the following $SL(N,\mathbb{C})$ gauge transformations:
\al{
\label{eq:slncgauge}
\bs
A_{\bar{z}}=&\hat U^{\dagger -1}(-\partial_{\bar{z}}H^{ }_2 H_2^{-1})\hat U^{\dagger}-\partial_{\bar{z}}\hat U^{\dagger -1}\hat U^{\dagger},\\
\mc{A}_z=&\hat U^{\dagger -1}(-\partial_z H^{ }_2 H_2^{-1})\hat U^{\dagger}-\partial_z\hat U^{\dagger -1}\hat U^{\dagger},\\
\tilde A_{\bar{z}}=& U^{\dagger -1}(-\partial_{\bar{z}}H^{ }_1 H_1^{-1}) U^{\dagger}-\partial_{\bar{z}} U^{\dagger -1}U^{\dagger},\\
\tilde{\mc{A}}_z=&U^{\dagger -1}(-\partial_z H^{ }_1 H_1^{-1}) U^{\dagger}-\partial_z U^{\dagger -1}U^{\dagger}.
\es
}
With using \eqref{eq:slncgauge}, $\mc{W}_R(C_1)$ and $\mc{T}_R(C_2)$ can be written as
\al{
\bs
\mc{W}_R(C_1,H_2)=Tr_R\ \mc{P}\ exp{\ \oint \limits_{C_1} (\partial_z H^{ }_2H_2^{-1}dz+\partial_{\bar{z}}H^{ }_2 H_2^{-1}d\bar{z})},\\
\mc{T}_R(C_2,H_1)=Tr_R\ \mc{P}\ exp{\ \oint \limits_{C_2} (\partial_z H^{ }_1 H_1^{-1}dz+\partial_{\bar{z}}H^{ }_1H_1^{-1}d\bar{z})}.
\es
}
The expectation value of the product of these two observables is given by
\al{
\label{eq:WTexval}
\bs
\langle \mc{W}_{R_1}(C_1)\mc{T}_{R_2}(C_2) \rangle=&\int d\mu(\mathscr{A}) \psi^*\mc{W}_{R_1}(C_1)\mc{T}_{R_2}(C_2)\psi\\
=&\int d\mu(H_1)d\mu(H_2)e^{\pa{2c_A+\frac{k}{2}}S_{WZW}(H_1)+\pa{2c_A-\frac{k}{2}}S_{WZW}(H_2)}\\
&\times \mc{W}_{R_1}(C_1,H_2)\mc{T}_{R_2}(C_2,H_1) + \mc{O}(1/m^2).
\es
}
For even level numbers, \eqref{eq:WTexval} can be written in terms of CS Wilson loop expectation values as
\al{
\label{eq:equivalence}
\langle \mc{W}_{R_1}(C_1)\mc{T}_{R_2}(C_2)\rangle_{YM_{2k}} = \bigg(\langle \mc{W}_{R_1}(C_1)\rangle_{CS_{k}}\bigg)\bigg(\langle \mc{W}_{R_2}(C_2)\rangle_{CS_{-k}}\bigg)+\mc{O}(1/m^2).
}
In the fundamental representation, the right-hand side of \eqref{eq:equivalence} is given by the HOMFLY polynomial\cite{freyd1985new,Witten:1988hf}. Thus, the left-hand side is indirectly given by skein relations, therefore it is a link invariant.

%------------------------------------------------------------------------------------------------------------------------

\section{Conclusion}

In this paper, we have studied geometric quantization of pure YM theory using holomorphic polarization. We have shown that three dimensional pure YM theory acts like a topological theory at certain limited scales. When the distance is large enough but finite, YM theory splits into two CS terms with levels $k/2$ and $-k/2$, very similar to the CS decomposition of the analogous TMAdSG model in its corresponding limit. At very large distances($m\rightarrow\infty$), these two CS terms cancel to make YM theory trivial, as required by the existence of a mass gap. Previous studies on geometric quantization of pure YM theory used real polarizations, which facilitates studying shorter distances. However, holomorphic polarization can probe into large distances, where CS splitting is observed. The reasoning is as follows. Holomorphic polarization makes the wave-functional explicitly $E$-field dependent, while real polarization makes it explicitly $B$-field dependent. At large distances $E$-fields dominate and only holomorphic polarization can resolve this scale. For shorter distances $B$-fields become dominant and this scale can be resolved by real polarization.

Together with our previous work on TMYM theory\cite{yildirim}, we have shown that both YM and TMYM theories exhibit CS splitting at large scales, just as predicted by their gravitational analogues. These results show that the analogy between TMYM theory and TMAdSG is more extensive than previously thought. For both YM and TMYM theories, this limited topological region can be exploited to incorporate link invariants of knot theory.

The CS splitting behavior appears in many problems in condensed matter physics and it is usually referred to as \emph{doubled Chern-Simons}. Quantum Spin Hall effect\cite{bernevig2006quantum} offers an example for CS splitting of the form \eqref{eq:YMlimit} at low energies. Our result may imply the existence of an underlying YM theory at higher energies. Similarly, a CS splitting of the form \eqref{eq:tmymlimit} at low energies may imply an underlying TMYM theory at higher energies.

\subsection*{Acknowledgment}
I am thankful to Vincent Rodgers and Parameswaran Nair for their supervision and support. 

%-------------------------------------------------------------------------------------------------------------------------
%-------------------------------------------------------------------------------------------------------------------------

%--------------------------------------------------------------------------------------
\begin{center}
\noindent\line(1,0){250}
\end{center}

\bibliographystyle{unsrt}

\footnotesize

\begin{thebibliography}{10}

\bibitem{achucarro1986chern}
A.~Ach{\'u}carro and P.K. Townsend.
\newblock {A Chern-Simons action for three-dimensional anti-de Sitter
  supergravity theories}.
\newblock {\em Physics Letters B}, 180(1):89--92, 1986.

\bibitem{witten1988}
E.~Witten.
\newblock {2+ 1 Dimensional Gravity As An Exactly Soluble System}.
\newblock {\em Nuclear Physics B}, 311(1):46--78, 1988.

\bibitem{Carlip2008272}
S.~Carlip, S.~Deser, A.~Waldron, and D.K. Wise.
\newblock {Topologically Massive AdS Gravity}.
\newblock {\em Physics Letters B}, 666(3):272 -- 276, 2008.

\bibitem{Carlip2}
S.~Carlip, S.~Deser, A.~Waldron, and D.K. Wise.
\newblock {Cosmological Topologically Massive Gravitons and Photons}.
\newblock {\em Classical and Quantum Gravity}, 26(7):075008, 2009.

\bibitem{Deser1982372}
S.~Deser, R.~Jackiw, and S.~Templeton.
\newblock {Topologically Massive Gauge Theories}.
\newblock {\em Annals of Physics}, 140:372 -- 411, 1982.

\bibitem{Deser2}
S.~Deser, R.~Jackiw, and S.~Templeton.
\newblock Three-dimensional massive gauge theories.
\newblock {\em Physical Review Letters}, 48:975--978, 1982.

\bibitem{yildirim}
T.~Yildirim.
\newblock {Topologically Massive Yang-Mills Theory and Link Invariants}.
\newblock {\em International Journal of Modern Physics A}, 30(7):1550034, 2015.

\bibitem{Bos:1989kn}
M.~Bos and V.P. Nair.
\newblock {Coherent State Quantization of Chern-Simons Theory}.
\newblock {\em International Journal of Modern Physics A}, A5:959, 1990.

\bibitem{Karabali1996135}
D.~Karabali and V.P. Nair.
\newblock {A Gauge-Invariant Hamiltonian Analysis for Non-Abelian Gauge
  Theories in (2+1) Dimensions}.
\newblock {\em Nuclear Physics B}, 464:135 -- 152, 1996.

\bibitem{Karabali:1996iu}
D.~Karabali and V.P. Nair.
\newblock {Gauge Invariance and Mass Gap in (2+1)-Dimensional Yang-Mills
  Theory}.
\newblock {\em International Journal of Modern Physics A}, A12:1161--1172,
  1997.

\bibitem{Karabali:1998yq}
D.~Karabali, C.-J. Kim, and V.P. Nair.
\newblock {On the Vacuum Wave Function and String Tension of Yang-Mills
  Theories In (2+1)-Dimensions}.
\newblock {\em Physics Letters B}, B434:103--109, 1998.

\bibitem{Karabali:1999ef}
D.~Karabali, C.-J. Kim, and V.P. Nair.
\newblock {Gauge Invariant Variables and the Yang-Mills-Chern-Simons Theory}.
\newblock {\em Nuclear Physics B}, B566:331--347, 2000.

\bibitem{Nair:2005iw}
V.P. Nair.
\newblock {\em {Quantum Field Theory: A Modern Perspective}}.
\newblock Springer, 2005.

\bibitem{hall2013}
B.~C. Hall.
\newblock {\em {Quantum Theory for Mathematicians}}.
\newblock Springer, 2013.

\bibitem{Polyakov1983121}
A.~Polyakov and P.B. Wiegmann.
\newblock {Theory of Nonabelian Goldstone Bosons In Two Dimensions}.
\newblock {\em Physics Letters B}, 131:121 -- 126, 1983.

\bibitem{Polyakov1984223}
A.M. Polyakov and P.B. Wiegmann.
\newblock {Goldstone Fields In Two Dimensions With Multivalued actions}.
\newblock {\em Physics Letters B}, 141:223 -- 228, 1984.

\bibitem{Gawedzki:1988hq}
K.~Gawedzki and A.~Kupiainen.
\newblock {G/h Conformal Field Theory from Gauged WZW Model}.
\newblock {\em Phys. Lett.}, B215:119--123, 1988.

\bibitem{Gawedzki:1988nj}
K.~Gawedzki and A.~Kupiainen.
\newblock {Coset Construction from Functional Integrals}.
\newblock {\em Nucl. Phys.}, B320:625--668, 1989.

\bibitem{Lemes:1998md}
V.E.R. Lemes, C.~Linhares de~Jesus, S.P. Sorella, L.C.Q. Vilar, and O.S.
  Ventura.
\newblock {Chern-Simons as a Geometrical Setup for Three-Dimensional Gauge
  Theories}.
\newblock {\em Physical Review D}, D58:045010, 1998.

\bibitem{Grignani1997360}
G.~Grignani, G.~Semenoff, P.~Sodano, and O.~Tirkkonen.
\newblock {G/G Models as the Strong Coupling Limit of Topologically Massive
  Gauge Theory}.
\newblock {\em Nuclear Physics B}, 489:360 -- 384, 1997.

\bibitem{freyd1985new}
P.~Freyd, D.~Yetter, J.~Hoste, W.B.R. Lickorish, K.~Millett, and A.~Ocneanu.
\newblock {A New Polynomial Invariant of Knots and Links}.
\newblock {\em Bulletin of the American Mathematical Society}, 12(2):239--246,
  1985.

\bibitem{Witten:1988hf}
E.~Witten.
\newblock {Quantum Field Theory and the Jones Polynomial}.
\newblock {\em Communications in Mathematical Physics}, 121:351, 1989.

\bibitem{bernevig2006quantum}
B~Andrei Bernevig and Shou-Cheng Zhang.
\newblock Quantum spin hall effect.
\newblock {\em Physical Review Letters}, 96(10):106802, 2006.

\end{thebibliography}
%--------------------------------------------------------------------------------------

\end{document}